\documentclass[fleqn,12pt,twoside]{article}
\usepackage{espcrc1}
\usepackage{graphicx}
\usepackage[figuresright]{rotating}

\newcommand{\AmS}{{\protect\the\textfont2
  A\kern-.1667em\lower.5ex\hbox{M}\kern-.125emS}}
\newcommand{\gsim}{\raisebox{-5pt}{$\;\stackrel{\textstyle >}{\sim}\;$}}

\newcommand{\beq}{\begin{eqnarray}}
\newcommand{\eeq}{\end{eqnarray}} 
\newcommand{\Gev}{{\rm GeV}}
\newcommand{\Mev}{{\rm MeV}}   
\newcommand{\eq}{\ref}    
 
\title{Gamma-Ray Imaging 
    by Silicon Detectors in Space: the AREM Method}

\author{Carlotta Pittori\address[TOV]
{Universit\`a di Roma  ``Tor Vergata'' and 
INFN, Sez. di Roma II, Via della Ricerca Scientifica 1, 
I-00133 Rome, Italy.}\thanks{Corresponding author. 
{\it E-mail addess:} carlotta.pittori@roma2.infn.it ; 
fax.: 39-06-7259-4647.}
        ~and
        Marco Tavani\address[IFC]{Istituto di Fisica Cosmica G. Occhialini 
- CNR, Milano, Italy.}
 }      
\input epsf
\begin{document}
\maketitle

\begin{abstract}
We present  the Agile REconstruction Method (AREM) for
$\gamma$-ray (30 MeV -- 50 GeV) direction
reconstruction applicable to high-resolution Silicon Tracker
detectors in space. It can be used in a ``fast mode'',
independently of Kalman filters
techniques, or in an ``optimized mode'', including
Kalman filter algorithms for track identification.
AREM correctly addresses three points of the analysis
which become relevant for
off-axis incidence angles:
(1) intrinsic ambiguity 
in the identification of the 3-dimensional 
$e^+/e^-$ tracks; (2) proper identification 
of the 3-dimensional pair production plane 
and reconstructed direction;   
(3) careful choice of an energy weighting scheme
for the 3-dimensional tracks.
We apply our method to simulated gamma-rays
in the AGILE detector. The excellent
spatial resolution obtained by the AGILE Silicon Tracker,
providing crucial analog information,
makes it possible to improve the
spatial resolution of previous detectors (e.g. EGRET) by 
a factor of $\sim 2$  in containment radius at $E_{\gamma} \gsim 400$ MeV.
In this
paper we present the results of our
3D-method for a selected sample of photon incidence angles and
energies. A more comprehensive and complete discussion including the
use of Kalman filter algorithms will be the subject of forthcoming
papers. 
\end{abstract}
\section{Introduction}
High energy gamma-ray astrophysics is foreseen to be 
one of the more challenging fields of study in the coming years.
Previous gamma-ray missions such as SAS-2 \cite{SAS2},
COS-B \cite{COSB}, and especially 
the recent EGRET experiment \cite{EGRET,1catalog,3catalog} 
on the Compton Gamma Ray Observatory, left us with 
a large amount of exciting results and open questions.
Indeed, of the nearly 300 gamma-ray sources detected so far
only a small fraction ($ \sim 30 \% $) has been identified \cite{3catalog}.
The discovery of gamma-ray blazars, pulsars, high-energy
gamma-ray bursts (GRBs) and of a large amount of unidentified 
sources, many of which are strong high-energy transients, 
has provided 
clear evidence for the need of  
a next generation of gamma-ray experiments
with increased field of view and improved angular resolution.
In this context, the AGILE mission \cite{AGILE}, 
planned to be operational during the year 2003,
is integrated toward the GLAST mission \cite{GLAST},
planned for the year 2006. 
In this paper, we consider
the EGRET spark chamber experiment 
as the direct AGILE predecessor, 
and we will refer to it for comparisons.
\par
We recall here that AGILE (Astro-rivelatore Gamma a Immagini 
LEggero)\footnote{
More information about the project can be found at the site:
\it{http://www.ifctr.mi.cnr.it/Agile}.
}
is a Small Scientific Mission of ASI (Agenzia Spaziale Italiana)
with a tracking system based on the state-of-the-art Silicon strip 
technology \cite{AGILE,barbiellini}.
The AGILE GRID (Gamma-Ray Imaging Detector) consists of a 
Silicon-Tungsten Tracker, a Cesium-Iodide Mini-Calorimeter
and a segmented Anticoincidence system of plastic scintillators
and is sensitive to 30 MeV - 50 GeV photons. 
Thanks to the fast readout electronics and to the segmented 
Anticoincidence system, AGILE will have, among other features,
an unprecedently large field of view (FOV $\sim 3$ sr),
 larger than previous gamma-ray experiments 
by a factor $\sim 5$.
Furthermore the Silicon Tracker has
a very good intrinsic spatial resolution,
comparable or even less than the microstrip 
pitch $ < 121 ~\mu$m, 
by using the analog information on the released
charge distribution between strips. 
For comparison, we recall that the 
EGRET spark chamber wire spacing was equal to
$820 ~\mu$m.
\par
The AGILE goal is to obtain the best sensitivity ever reached for 
off-axis events (up to $\sim 60^{\circ}$), and an
on-axis sensitivity 
comparable to that of EGRET, despite the 
smaller dimensions and effective area.
Therefore, the optimization of the angular resolution algorithms
becomes a crucial point to fulfil
the mission scientific objectives.
 \par
The $\gamma$-ray photon direction reconstruction is based on the physical
process of pair production, and is obtained from 
the identification and the detailed 
analysis of the electron/positron tracks stemming
from a common vertex.
The direction reconstruction should 
take into account the 
effect of multiple Coulomb scattering
and the distribution of the total 
energy of the incident photon
between the $e^+/e^-$ particles. 
Until now, this was done by analyzing separately the
two tracks projections in the ZX and ZY views.
We will show in section 2 that the ``2-D projection method''
is a good approximation only for nearly on-axis events,
but it induces two kinds of systematic error in the 
$\gamma$-ray
direction reconstruction for off-axis events because of
the intrinsic ambiguity in the proper identification of 
3-D tracks and of 
the problem of the correct reconstruction of the 
true 3-D photon direction.
 Finally, we point out the importance of the
choice of a proper track weighting scheme.
\par
In this paper we emphasize that,
contrary to previous $\gamma$-ray experiments,
the simplified 2-D projection method 
would not be a good approximation for AGILE
because of its large field of view and very good 
intrinsic spatial resolution. As we discuss in the next section,
the first source of error follows from the ambiguity 
in the association of coordinate pairs
$(X,Y)$ corresponding to the hits in the Tracker 
in the two orthogonal ZX and ZY views.
This ambiguity can be resolved 
in all those cases in which one of the two tracks becomes
distinguishable from the other in both 
views of the Tracker, 
as for example when it stops or exits the Tracker, or when
it suffers a more significant multiple scattering.
Regarding the second point, we notice that
in general, and expecially 
for off-axis events, the 3-D 
reconstruction 
is not equivalent to the composition of 
reconstructed directions
in each projected view. So 
we are forced to consider an intrinsically
three-dimensional strategy in order to  
reconstruct the true incident photon direction. 
The third point keeps into account the fact that,
in general, the photon energy is not evenly divided  
between the two pair particles. 
It turns out that the incident photon
direction is closer to that of the most energetic particle, 
and an ``energy-weighted'' direction
reconstruction is necessary.  
\par
The AGILE REconstruction Method (AREM)
presented in this paper, provides a general baseline,
to be optimized for each particular 
$\gamma$-ray instrument, 
to keep into account these three points of the 
analysis. 
\section{The AREM method}
\label{sect:method}

\subsection{The conversion plane problem}
\label{sub:error1}
\begin{figure}
\hspace*{4cm}
\epsfxsize=7 cm  \epsfbox{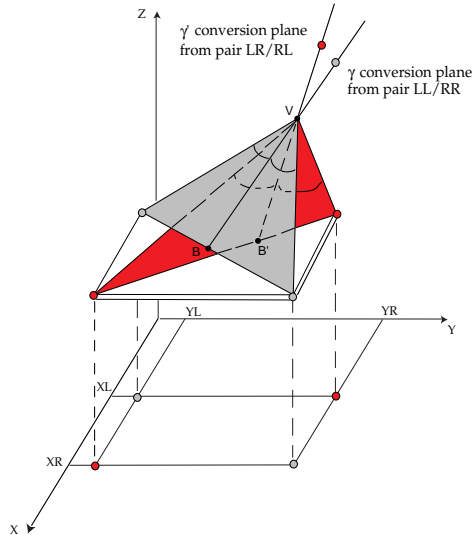} 
\vspace*{-1 cm}
\caption{The ``conversion plane problem'':
a correct event reconstruction in 3-D implies
solving the coordinate track ambiguity and making the 
right choice for one out of two possible conversion
planes. 
}
\vspace*{-0.5cm}
\label{fig:tria}
\end{figure}

Let us first 
illustrate and discuss the conversion plane problem
in the simplified case of an even energy sharing
between the electrons and positrons 
i.e., when the photon direction coincides with the bisector.
We restrict the discussion
to the idealized case of 
absence of 
$\delta$-rays or other secondaries 
in the Tracker.
In general, each Si-Tracker plane of a 
$\gamma$-ray detector 
is composed of a converter layer
and by two orthogonal 
strips layers.
Charged particles entering the Tracker will 
produce an electric signal in the front-end electronics
which would correspond to the hit readout
strips in the two 
active X and Y layers. As long as a single charged particle
crosses the Tracker, this signal will identify a 
unique point in 3-D space for each hit plane.
However, when two separated particles
hit simultaneously the active layers (as in the case with
$e^+/e^-$ pairs) 
the signal could correspond to two possible couples
of points in 3-D space, as shown in Fig. \ref{fig:tria}. 
The two possible couples
of points are:
pair LL/RR = $\left[(X_L,Y_L),(X_R,Y_R)\right]$, open symbols
in the figure, 
{\bf or}
pair LR/RL =
$\left[(X_L,Y_R),(X_R,Y_L)\right]$, filled symbols. 
This gives rise to an intrinsic ambiguity in the 
conversion plane identification.
In terms of projected views this ``conversion plane problem''
can be phrased as:
``Does the track to the left in the ZX view correspond 
to the 
left track in the ZY view - pair LL/RR? Or does it correspond to the
track to the right - pair LR/RL?''. 
\begin{figure} [tbh]
\hspace*{4cm}
\epsfxsize=8 cm \epsfbox{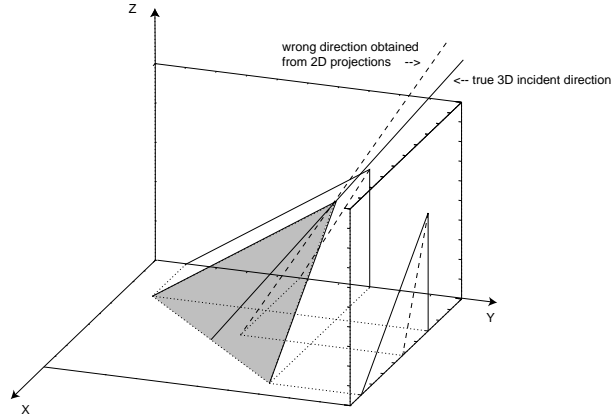} 
\caption{``The projection problem''
for the idealized case of an even energy share between pair
particles: the true 3-D bisector 
(solid line in the shaded
conversion plane) is different from the one obtained from the 
two bisecting lines in each projected view
(dashed line from dashed projections).}
\label{fig:bisector}
\end{figure}
Note that
for on-axis events the ``pyramid'' in 
Fig. \ref{fig:tria} becomes 
isosceles and 
the conversion plane problem apparently 
disappears, since both alternatives - pair LL/RR or LR/RL - 
would lead to the same bisector.
However, we recall that 
we are considering here the particular and unrealistic
case of an even energy sharing.  
In general
the photon energy is not evenly divided  
between the two particles (see Sect. \ref{sub:weight}), 
and the photon direction does not coincide with the bisector.
Hence the ambiguity 
for the identification of the ``energy-weighted''  
reconstructed direction applies also to on-axis events. \par
AREM algorithms solve the intrinsic ambiguity 
in the identification of the 3-D 
$e^+/e^-$ tracks by identifying
a primary and a secondary track in both ZX and ZY views.
The primary track is defined as 
that one carrying 
most of the incident energy, 
and subject to less multiple scattering. 
In practice, we identify
the primary track as the one for which
(starting from the third plane)
the quantity $|X_{CLUSTER}-X_{EXTRAPOLATED}|$
is on average minimized.
This means that the primary track 
should have the smallest
mean distance between the selected clusters 
and the position obtained by 
extrapolation of the trajectory
from the previous planes.
The 3-D primary track is then obtained by associating
the primary track in the ZX view with the 
primary track in the ZY view
(analogously for the secondary one).
\subsection{The projection problem}
\label{sub:error2}
According to the simplified 2-D projection method,
the first step is to identify
the two projected tracks in each view. The following step
is to take their 
(in general weighted) bisectors and compose
them to obtain the reconstructed gamma direction.
This procedure is not correct: 
the true 3-D bisector
has ZY and ZY projections which {\bf do not} correspond
to the two bisecting lines in each view. This is illustrated in 
Fig. \ref{fig:bisector}.
The only case in which this equality holds true is for on-axis events.
As shown in ref. \cite{andrea}, this systematic effect 
increases for increasing off-axis angles and large opening angles
up to values of $\sim 0.5^{\circ}$, hence the 2-D projection method 
is not acceptable for AGILE and similar 
detectors\footnote
{We warmly thank the EGRET collaboration,
and in particular D.L. Bertsch and D. Thompson, for allowing us to perform a
test of our reconstruction algorithms
on their calibration data.
In the case of EGRET data
this systematic error 
can be considered almost negligible, since it is hidden
by the relatively low spark chamber intrinsic resolution.}.
\par
The proper identification with AREM
of the 3-D pair production plane and 
photon direction reconstruction 
can be obtained by several completely equivalent
geometrical methods, such as: spherical triangle formulae, 
normalized 3-D vector components, or double
rotation of reference system.
In the following we use the latter method.
\subsection{Track weighting scheme}
\label{sub:weight}
In the pair production
process the total energy of the incident photon
is split between the $e^+/e^-$ particles. We recall 
that the probability distribution 
of having an electron, or a positron, with energy 
$E_{e^+(e^-)}$
when the primary photon has energy $E_{\gamma}$,
is almost flat 
in the ``low'' energy range $E_{\gamma}=10 \div 40$ MeV.
However, for increasing incident energy 
it is more and more
probable that one of the two 
secondary particles carries most 
of the total energy (see for example
Fig. 2.19.2 of ref. \cite{brossi}).
The RMS emission angle $\alpha_{e^+(e^-)}$ 
between the original photon direction and
the direction of the electron, or positron,
roughly depends on the inverse of the photon energy \cite{brossi}:
$\langle \alpha_{e^+(e^-)} \rangle \propto 1/E_{\gamma}$.
This implies that for most events 
in the AGILE energy range 
the main information on the original photon 
direction will be carried by
the most energetic $e^+(e^-)$ particle.
It follows that the choice of a proper track weighting 
scheme to keep into account
the energy distribution  
between the $e^+/e^-$ particles plays a fundamental r\^ole
in the angular reconstruction algorithms.
\par
Can we obtain a satisfactory track energy estimate
relying mainly on Tracker data?
Multiple scattering 
spreads the information on the 
original particle directions, but on the other side 
it can be exploited to obtain a track energy estimate.
According to the Moli\`ere theory \cite{moliere}, the 
small planar deflection angle of multiple scattering
$\beta$ 
has an approximately Gaussian distribution with high tails 
(for a detailed analysis 
see also ref. \cite{bert}).
The Moli\`ere formula for the RMS scattering angle can be written as:
\beq
\langle \beta \rangle = \frac{13.6 ~\Mev~}{pv}
\sqrt{\frac{X}{X_0}} \left[1 + 0.038 \ln(\frac{X}{X_0})\right],
\label{eq:moliere}
\eeq
where $X/X_0$ is the thickness of the scattering medium in 
radiation lengths. 
In our case, the relation between particle 
momentum and energy is: $pc \simeq E_{e^+(e^-)}$,
and the velocity is such that: $v/c \simeq 1$,
since we consider $E_{e^+(e^-)} >> m_e c^2$.
There is then an (approximate) inverse relation between
particle energy and RMS deflection angle: 
\beq
E_{e^+(e^-)} \propto \frac{ \sqrt{cos~\theta}}
{\langle \beta \rangle},
\label{eq:moliere2}
\eeq
where here $\theta$ is the angle 
between the 
considered track and the vertical axis. 
This means that the most energetic particle 
suffers less multiple Coulomb scattering.
\par
The AREM choice of weighting scheme
is based on Moli\`ere multiple scattering theory. 
In principle, 
each 3-D track should be weighted by
its energy to some power.
According to eq. (\eq{eq:moliere2}), 
for each projected track we define a weight as:
\beq
w ~\propto ~ (E_{e^+(e^-)})^r \propto (\sqrt{cos(\theta)}
/\langle \beta \rangle )^r,
\label{eq:pesi}
\eeq
where $r$ is a parameter
to be optimized for each particular 
$\gamma$-ray instrument. In practice, 
the quantity $|(X_{CLUSTER}-X_{EXTRAPOLATED})|$
is used to estimate the RMS multiple scattering
angles $<\beta>$, and hence the particle energies $E_{e^+(e^-)}$.
\subsection{The AREM flow}
\label{sub:flow}
\begin{figure}
\hspace*{3.5cm}
\epsfxsize=8 cm \epsfbox{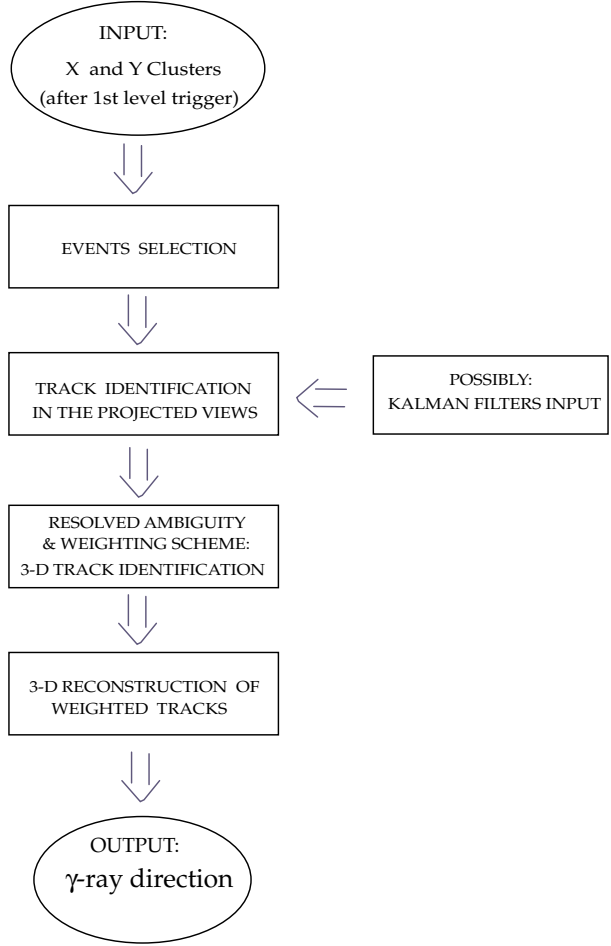} 
\caption{Flow chart
for the software corresponding to the AREM method.}
\label{fig:arem}
\end{figure}
Summing up, we show in Fig. \ref{fig:arem} the general flow chart
for the software corresponding to the AREM method described in
the previous subsections.
\section{Applications to the AGILE Tracker}
\subsection{Instrumental and geometrical set-up} 
\label{sub:setup}
The AGILE baseline configuration \cite{AGILE} 
is the following:
\begin{itemize}
\vskip -0.3 truecm
\item[-] Silicon Tracker:
the AGILE Tracker consists of 14 detection planes 
$42 \times 42$ cm$^2$ total area
with $1.6$ cm interplane distance.
Each plane is composed by two layers of 
microstrip Silicon detectors
with orthogonal strips (X and Y view). 
The first 12 planes, starting from the top, 
have also a Tungsten conversion layer each,
with thickness equal to $0.07 ~X_0$ radiation length.
The last two planes have no Tungsten layers,
since the readout trigger requires at least
three Si-layers 
providing signals.
The active area of the Si-detectors 
is made of $4 \times 4$ tiles 
where $4 \times 768$ microstrips 
of pitch of $121 ~{\mu}\rm{m}$ are implanted.
Only half of the strips in every tile are readout strips. 
However,
it is possible to detect a signal even if the 
particle hits a ``floating'' strip,
by taking into account the effect
of the capacitive coupling between contiguous strips\footnote
{The ``floating strip'' configuration has been chosen to 
achieve an excellent spatial resolution while
keeping
under control the number of readout channels and hence the detector
power consumption.}.
\item[-] Mini Calorimeter:
it consists of two orthogonal
planes, each containing 16 CsI bars,
for a total radiation length of $\sim 1.5 ~X_0$.
\item[-] Anticoincidence System: it is made of
lateral segmented Plastic Scintillator planes composed of
3 panels for each Tracker side and of a top plane 
of thickness of $\sim 0.5$ cm. 
\end{itemize}
\subsection{Monte Carlo simulations}
The Monte Carlo simulations of AGILE performances
are done with the GEANT 3.21 code \cite{GEANT}.
The GEANT package reproduces all possible interactions
of radiation and particles through matter, such as
pair production, multiple scattering, Compton scattering,
etc.  \par
In the case of the AGILE Tracker,
CERN testbeams indicate that 
the experimental uncertainty in
the identification of 
the position of a cluster of adjacent strips
providing signal on a track
can be reduced at the level of $\sim 40 ~\mu$m
\cite{barbiellini}.
This result is achieved by using the
analog readout which gives information on the 
charge distribution released in the 
Si-microstrips,
and the floating strip configuration.
However, since the experimental validation of the parametrization of 
the capacitive coupling effect at 
large off-axis angles is still in progress,
in order to implement and test the AREM method
we use here a simplified version of the 
AGILE simulation code, originally developed
by the AGILE simulation software group \cite{VERO}.
In this simplified model,
which does not yet include
capacitive coupling and floating strips, 
all the strips are readout strips
and the digitization uncertainty is simply
determined by the pitch. 
This detector model is 
very accurate and 
detailed enough to
be fully suitable for our analysis.
\par
The simulated events are characterized
by the incident photon energy, $E$, and by
a fixed incident direction ($\theta$ , $\phi$)
with respect to the detector sistem of reference.
The algorithms which generate flux sources 
originating from a fixed direction
are optimized to correctly simulate 
a plane wave front from an infinite distance \cite{VERO}.
\subsection{Analysis of simulated events}
From our Monte Carlo simulations, we obtain
the complete
detector response to the incident $\gamma$-ray flux,
at several energies and incident angles.
We recall that the standard data analysis based on
Kalman filter algorithms \cite{kalma},
which make use of the  
information from all hit planes,
is optimized for 
charged particle track identification, 
but not specifically for $\gamma$-ray
direction reconstruction.
Hence, 
as a first step to test
our reconstruction method,
we perform an analysis of the 
first n-Planes Resolution, i.e. using only information 
from the first hit planes, 
for which the information 
on the original photon direction
is less 
disturbed
by multiple scattering.
In this preliminary study we 
present in particular our results 
regarding AREM 2PR and 3PR
(2PR = two planes resolution, 3PR = three planes resolution). 
\par
In view of our introductory discussion,
multiple scattering has a two-fold effect:
it 
disperses the information 
on the original direction of the particle going
through subsequent converter layers; 
on the other hand, it can be exploited 
to obtain a particle energy estimate.
Integrating AREM with
the high efficiency algorithms based on  
the Kalman filters 
will be the next step to improve the energy determination  
and weighting scheme,
and to further optimize the AGILE angular resolution
\cite{noi}.
\subsection{Event selection}
According to AGILE Level-1 trigger conditions,
there must be at least three
consecutive Tracker planes which give a signal
in both views\footnote{
The actual on-board first level trigger
is given by the coincidence of 3 out of 
4 consecutive Tracker planes,
to take into account the possibility of
one-plane failure.
Simulations show that
the final difference
on the total number of rejected events 
at subsequent trigger level is
of order of $1\%$ \cite{VERO}.}.
With respect to trigger Level-1,
we perform a further event selection by asking that
accepted events show only one cluster of strips on the first 
hit Tracker plane in both views.
This amounts to require that it be
possible to identify a ``clean'' conversion vertex
in both Tracker views.
We also require no more than two clusters
on the second hit Tracker plane, always in
both ZX and ZY views. In such a way we reject all events
with spurious hits on the second plane,
i.e., we require a ``clean'' second plane.
We do not impose any further
conditions on the third and 
subsequent planes.
\subsection{Two plane resolution}
We have analized first the so-called 
two plane resolution (2PR) of the Tracker, which 
takes into account only the information
coming from the first two hit planes.
The 2PR gives no further information on
the deviation from the initial track
direction obtained from the firts two planes
and we are
forced to  neglect
multiple scattering and energy distribution effects.
We can only take a random choice regarding 
the conversion plane problem
described in Sect. \ref{sub:error1},
and we get a coordinate ambiguity error
for about $50\%$ of the photons.
Regarding the projection problem described
in Sect. \ref{sub:error2},
we can still utilize the 3-D method 
to guarantee the proper reconstruction of the bisector.
\par
\subsection{Three plane resolution}
We then include in our analysis
the information coming from the third hit plane.
For the three plane resolution (3PR), the 
event selection criteria 
remain the same as in the 2PR case, and 
are the ones described above.
This implies that the efficiency of the 
two reconstruction algorithms are equal.
In this case we are in the condition to
follow the complete AREM flow
and to correctly address the intrinsic track ambiguity
and the projection problem.
Having identified a primary and a secondary track,
we can assign a weight to each track 
as described in Sect. \ref{sect:method}.
In the following, 
we chose as a particular weighting scheme:
$w \propto E_{e^+(e^-)}$, i.e. we set $r=1$ in
eq. (\eq{eq:pesi}),
and the track energy is estimated from 
the inverse of square root of the 
variance of the track linear fit.
\subsection{Results}
\begin{figure} 
\vspace*{-1.3cm}
\hspace*{0.5cm}
\epsfxsize=15cm \epsfbox{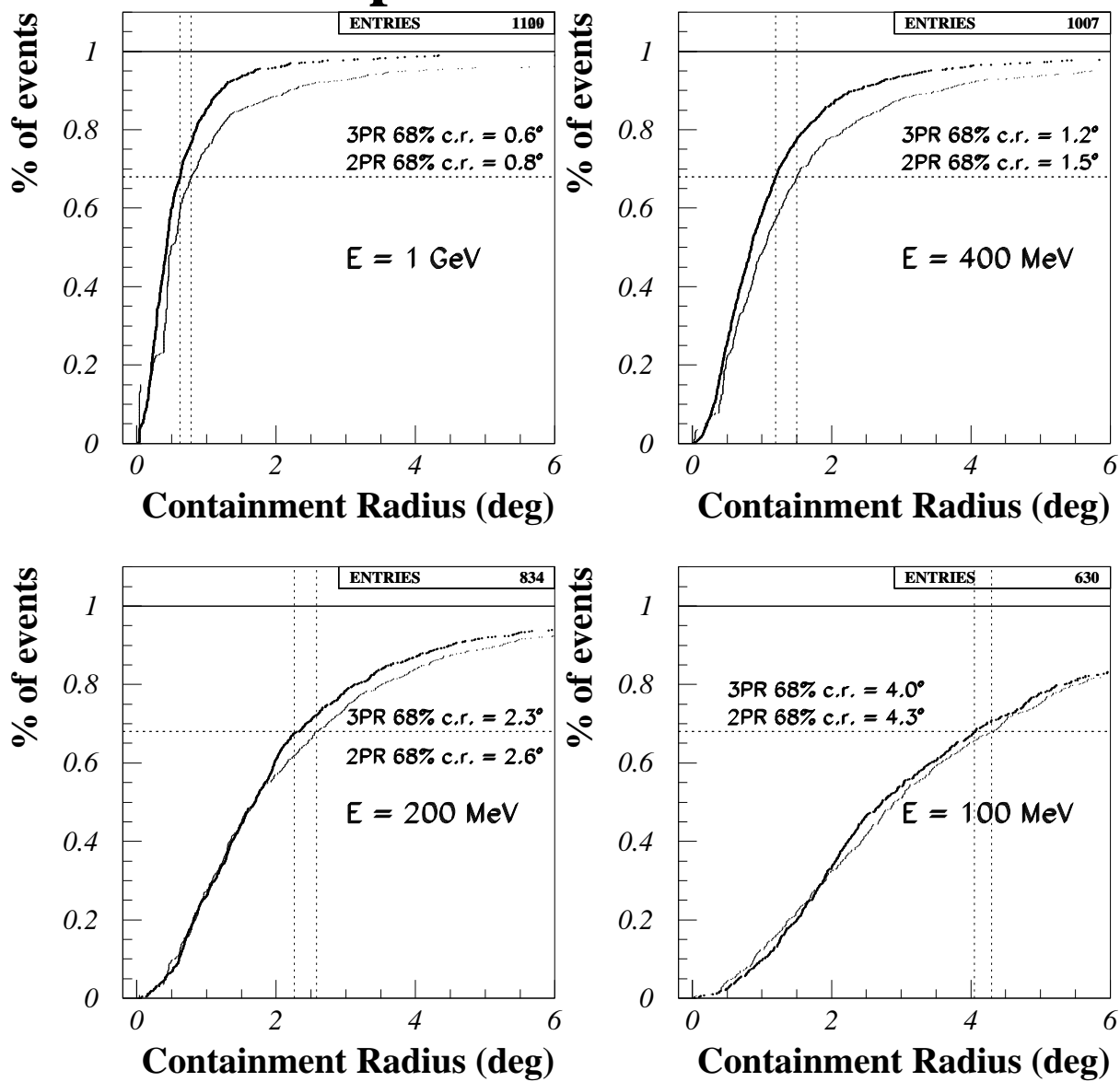} 
\vspace*{-0.5cm}
\caption{3-D AGILE PSF integral profiles, obtained with the
AREM $\rm{2PR}$ (thin dots) and $\rm{3PR}$ (thick dots)
for near on-axis events ($\theta=10^{\circ}$) at 
four incident energy values.
The intersection of the curves with the
horizontal dashed line corresponds to the PSF value
at 68 $\%$ c.r.}
\vspace*{-0.5cm}
\label{fig:a}
\end{figure}

\begin{figure}
\vspace*{-1.3cm}
\hspace*{0.5cm}
\epsfxsize=14cm \epsfbox{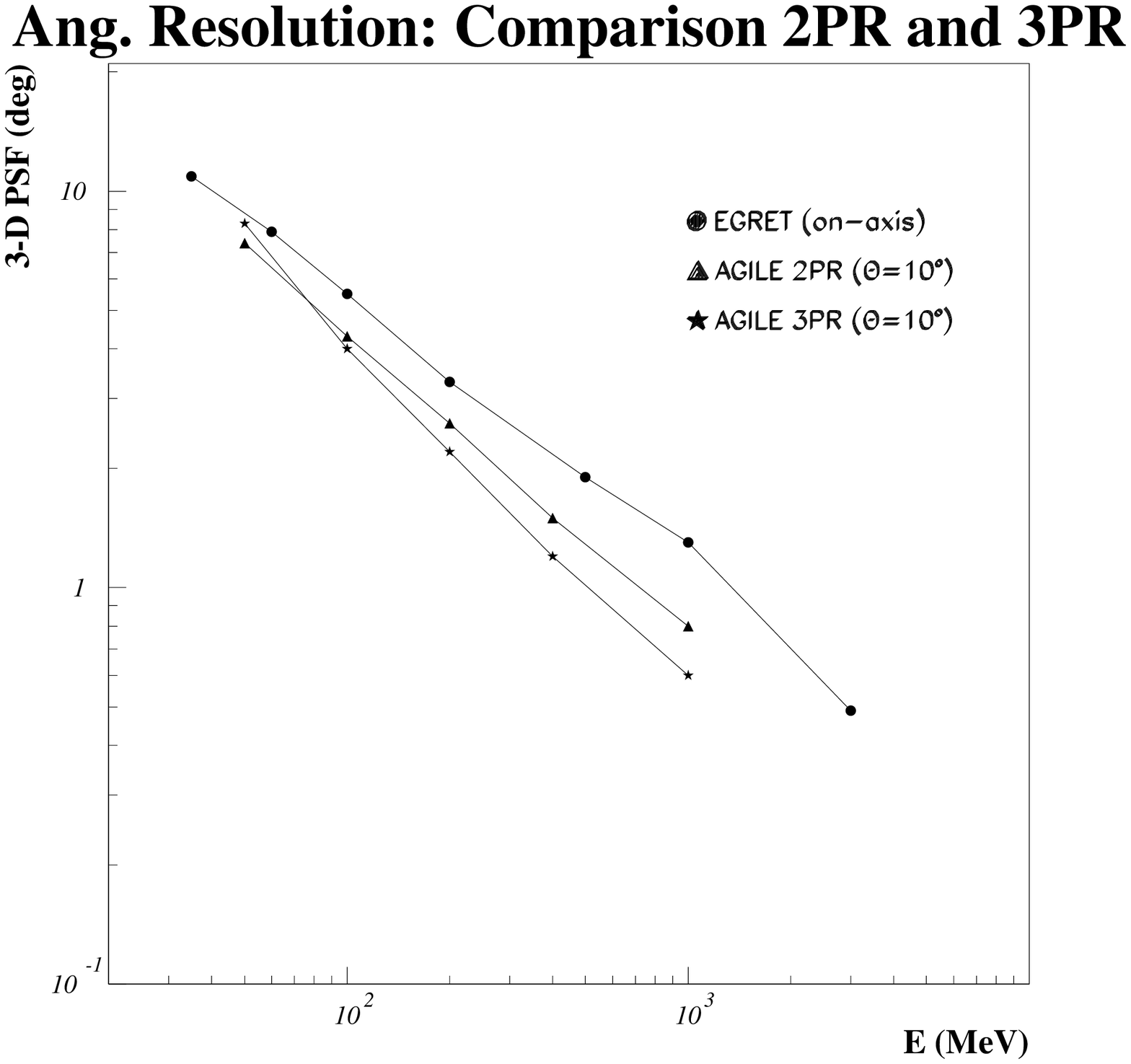} 
\vspace*{-0.5cm}
\caption{Comparison of 3-D PSF (at $68\%$ c.r.)
as a function of energy
in the case of EGRET (on-axis)
and of AGILE 2PR and 3PR (incidence angle $\theta=10^{\circ}$).}
\vspace*{-0.5cm}
\label{fig:b}
\end{figure}

In Fig. \ref{fig:a} 
we show the 3-D integral
Point Spread Function (PSF) 
for events at $\theta=10^{\circ}$, for the energies
$E=1 ~\Gev$, $E=400 ~\Mev$,
$E=200 ~\Mev$ and $E=100 ~\Mev$.
In that figure we compare the PSF profiles obtained by 
2PR and 3PR.
The 3PR $68\%$ containment radius (c.r.)
is systematically better than the 2PR one for 
$E \gsim 100 ~\Mev$.
Since the efficiency of the 
two reconstruction algorithms is the same,
the improvement of the angular resolution 
obtained with AREM 3PR 
is in part due to 
the solution of the coordinate track ambiguity
and of the adopted weighting scheme.
For the event set considered in Fig. \ref{fig:a}
this effect ranges from $\sim 10 \%$ up to $30\%$
at higher energy.
This is also shown in Fig. \ref{fig:b},  
were we
compare the 3-D PSF at $68\%$ c.r.
as a function of energy
in the case of EGRET on-axis
and of AGILE 2PR and 3PR at 
incidence angle $\theta=10^{\circ}$.
\par
\begin{figure}
\vspace*{-1.3cm}
\hspace*{0.5cm}
\epsfxsize=15cm \epsfbox{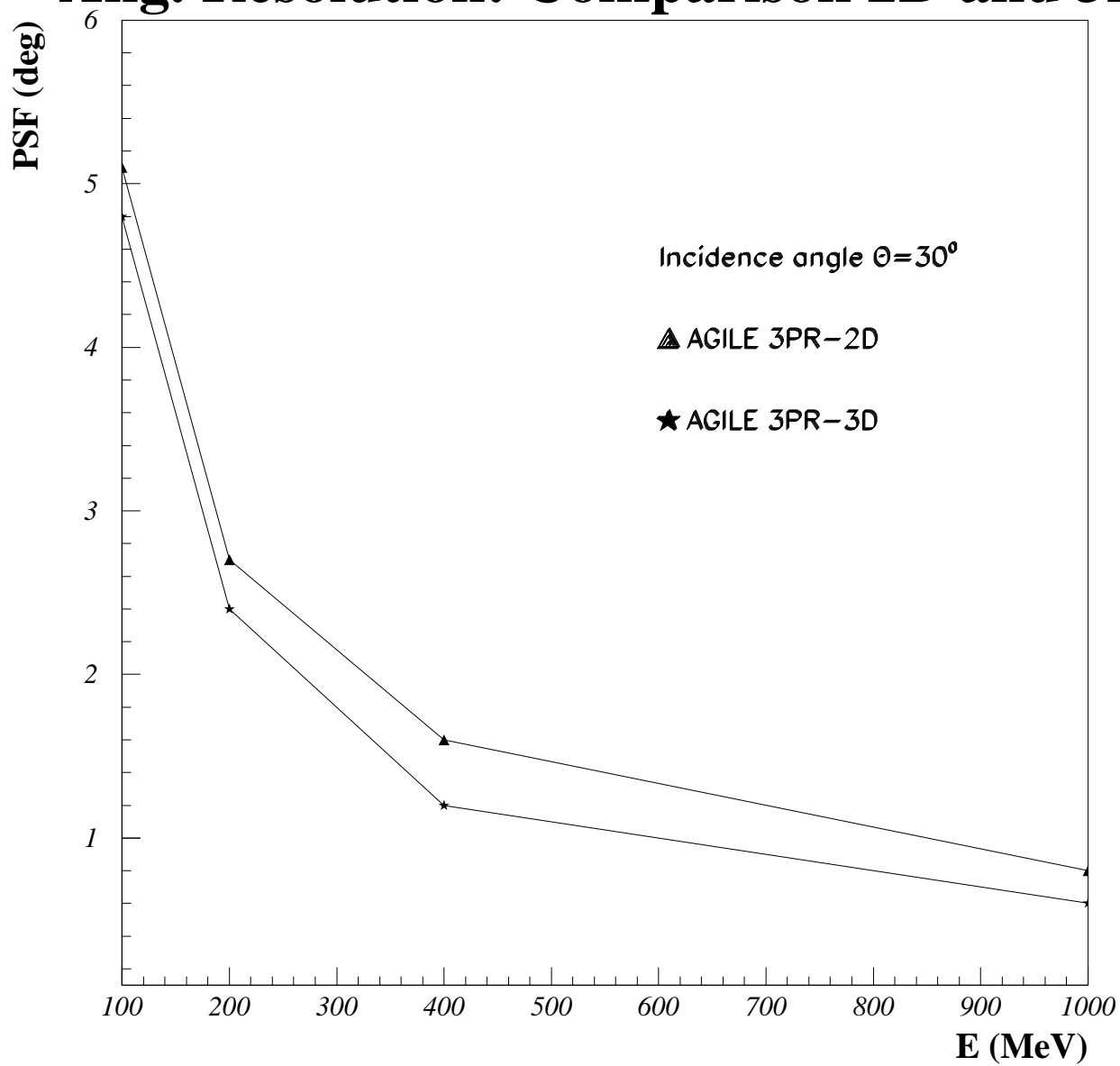} 
\vspace*{-0.5cm}
\caption{Comparison between the 2-D 
and 3-D reconstruction methods:
3PR PSF as a function of energy, 
at incidence angle $\theta=30^{\circ}$.}
\vspace*{-0.5cm}
\label{fig:2D3D}
\end{figure}

 
The comparison between the results obtained with the
2-D and 3-D reconstruction methods, both applied to the
3PR algorithm, is shown if Fig. \ref{fig:2D3D} 
which summarizes
the effect of AREM on the AGILE resolution at
incidence angle $\theta=30^{\circ}$.
\par
\begin{figure} 
\vspace*{-1.3cm}
\hspace*{0.5cm}
\epsfxsize=15cm \epsfbox{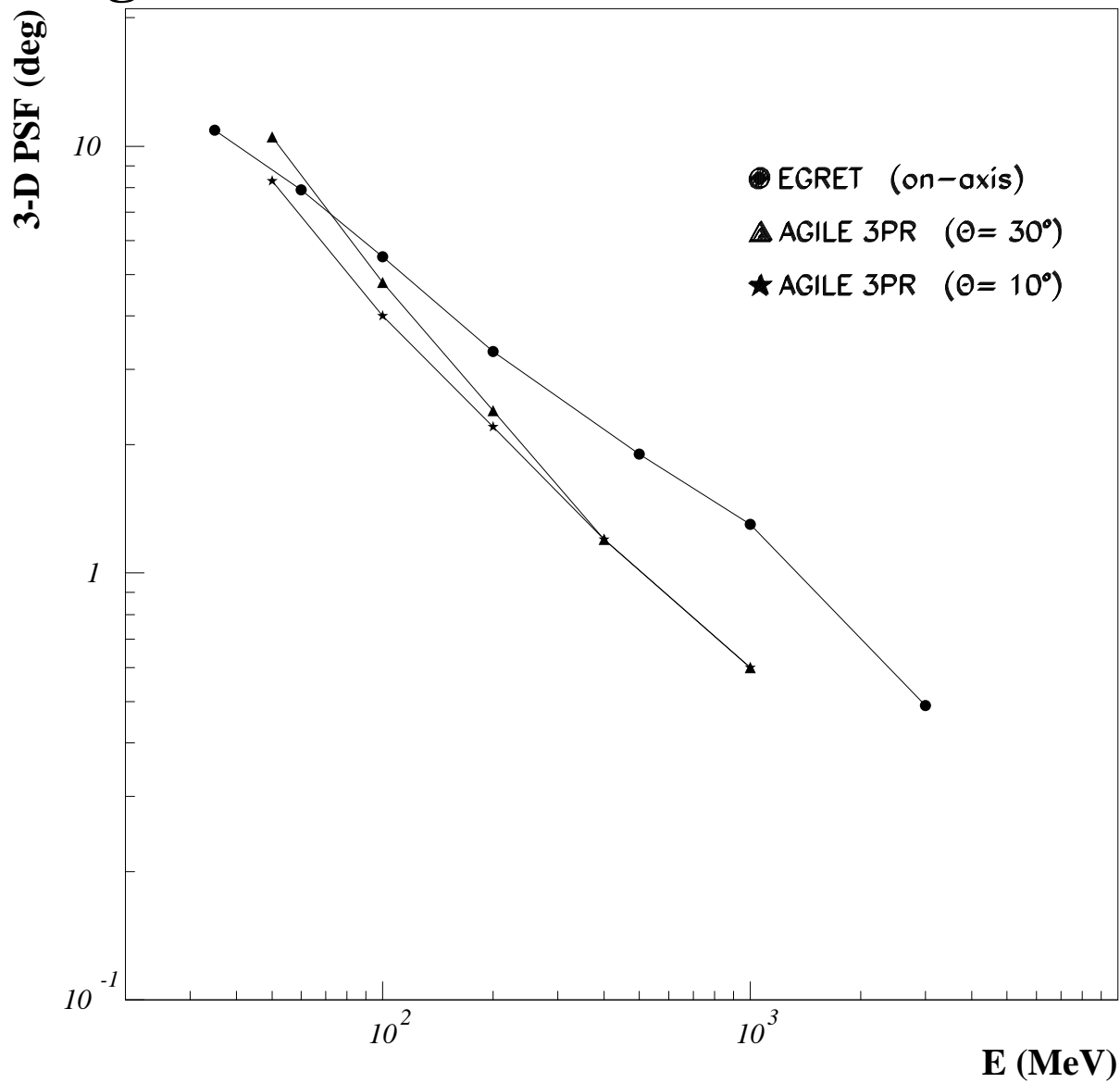} 
\vspace*{-0.5cm}
\caption{3-D PSF (at $68\%$ c.r.)
as a function of energy
obtained with AREM 3PR for 
AGILE, at incidence angles
$\theta=10^{\circ}$ and $30^{\circ}$.
EGRET on-axis resolution is also shown in the Figure.}
\vspace*{-0.5cm}
\label{fig:d}
\end{figure}
\begin{figure} 
\vspace*{-1.3cm}
\hspace*{0.5cm}
\epsfxsize=15cm \epsfbox{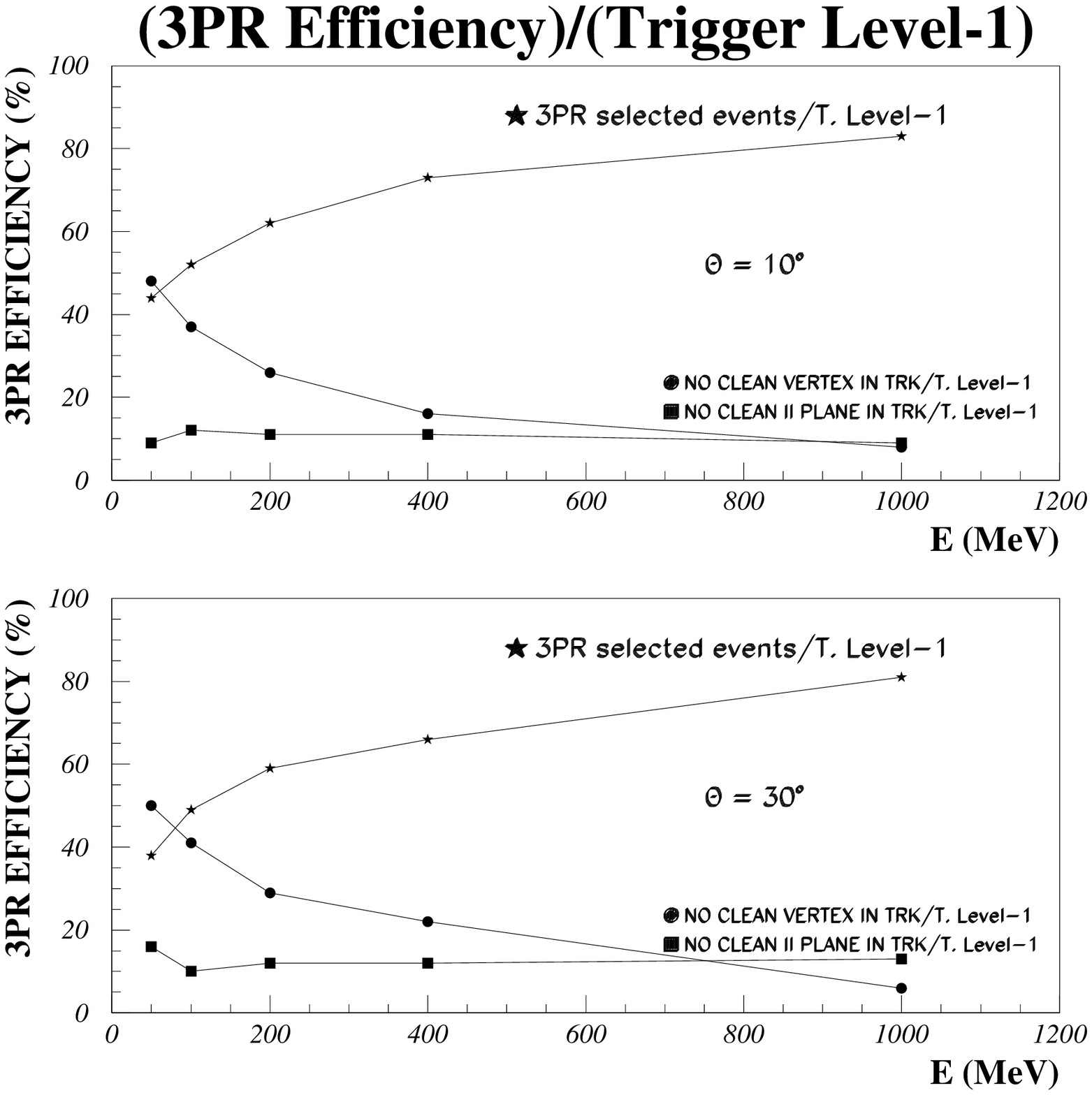} 
\vspace*{-0.5cm}
\caption{Efficiency of the AREM $\rm{3PR}$ method as a function of 
energy at incidence angle 
$\theta=10^{\circ}$ (upper panel) and $\theta=30^{\circ}$
(lower panel).}
\vspace*{-0.5cm}
\label{fig:c}
\end{figure}

In Fig. \ref{fig:d} we report the 
final results obtained with AREM 3PR for AGILE  
angular resolution near on-axis and
at $\theta=30^{\circ}$ off-axis.  
The efficiency of the AREM 3PR 
reconstruction (same as 2PR) 
is shown in Fig. \ref{fig:c}
and it is compatible with the one obtained by EGRET \cite{EGRET}.
Summing up, we find that 
the AGILE 3PR resolution is better than that of EGRET by
a factor of $\sim 2$ above 400 MeV.
These results are
remarkable considering
the fact that in this preliminary study
we are disregarding all the
information on the following hit planes, and that we make no use
yet of the Mini-Calorimeter information. 
\section{Conclusions}

We presented here the Agile REcontruction Method and
the results of a preliminary study of its application
using only information 
from the first three hit planes, for which
the information on the original photon direction  
is less 
dispersed
by multiple scattering.
The preliminary results of AREM
3PR provide a satisfactory PSF, that 
turns out to be
better than that of EGRET by
a factor of $\sim 2$ at energies above
or equal $400$ MeV,
up to incidence angles of $\theta=30^{\circ}$ off-axis.  
We expect to improve this result 
by integrating the AREM method with the 
standard Kalman filter algorithms,
to take properly into account the
information from all hit planes, 
and with the Mini Calorimeter data,
in order to optimize the track energy determination  
and the AREM weighting scheme \cite{noi}.
Moreover, a further
improvement of the AGILE PSF
should be obtained 
by applying the AREM method to 
a model of the AGILE instrument which
includes the forthcoming results of a more accurate 
experimental study of the charge deposition in the Si-microstrips.
This analysis 
will allow an optimal use of the analog readout, which is one of the
main characteristics of the AGILE Tracker.
\par
\section*{ACKNOWLEDGMENTS}     
We thank D.L. Bertsch and D. Thompson for many discussions
and collaborative work on EGRET calibration data.
We also thank V. Cocco, A. Giuliani, F. Longo,
S. Mereghetti, A. Pellizzoni,
S. Vercellone and D. Zanello for many interesting discussions
and scientific exchanges. We wish 
to thank the Astroparticle Physics WIZARD Group
and INFN Section of Roma II for kind hospitality.
C.P. thanks in particular M.P. De Pascale
for useful comments after a careful reading of the manuscript,
A. Morselli for collaboration at the initial stage of this work
and  P. Picozza for having set up the Cosmic-Ray research area at
the `` Tor Vergata'' University of Rome.
\par \noindent
Work carried out under the auspices of the Agenzia Spaziale
Italiana.

%
%
%

\begin{thebibliography}{10}
\bibitem{SAS2} C.E. Fichtel et al., Astroph. J. 198 (1975) 163. 
\bibitem{COSB} B.N. Swanenburg et al., Astroph. J. 243 (1981) L69. 
\bibitem{EGRET} D.J. Thompson et al., Astroph. J. Suppl. 86 (1993) 629. 
\bibitem{1catalog}  C.E. Fichtel et al., Astroph. J. Suppl. 94 (1994) 551.
\bibitem{3catalog} R.C. Hartman, et al., Astroph. J. Suppl. 123 (1999) 79.
\bibitem{AGILE} M. Tavani et al., Proceedings of
the GAMMA 2001 Symposium, to be published by AIP, and preprint (2001)
in preparation, to be submitted to Nucl. Instr. and Meth. A. 
\bibitem{GLAST} P. Michelson, Proceedings of
the GAMMA 2001 Symposium, to be published by AIP.  
\bibitem{barbiellini} Barbiellini, G. et al., Proceedings of
the GAMMA 2001 Symposium, to be published by AIP, and preprint (2001)
submitted to Nucl. Instr. and Meth. A. 
\bibitem{andrea} A. Giuliani, 
``Studio e Ottimizzazione della Risoluzione Angolare del
Telescopio Spaziale per Astronomia Gamma AGILE'',
Laurea Dissertation, Universit\`a degli Studi di Pavia, (2001).
\bibitem{brossi} B. Rossi,
{\it `` High-Energy Particles''}, Prentice-Hall, New York (1952).
\bibitem{moliere} G. Moli\`ere, Z. Naturforsch. 2a (1947) 133 and
3a (1948) 78.
\bibitem{bert} D.L. Bertsch, Nucl. Instr. and Meth. 220 (1984) 489. 
\bibitem{VERO} V. Cocco, F. Longo and M. Tavani,  
AGILE Internal Tech. Note, AGILE-SIM-TN-001, 
Issue n.2 (2000), and preprint (2001), to be published in
Nucl. Instr. and Meth. A. 
\bibitem{GEANT} GEANT {\it Detector Description and Simulation Tool}, 
1993, CERN.
\bibitem{kalma}  R. Fr\"uhwirth, Nucl. Instr. and Meth. A 262 (1987) 444.  
\bibitem{noi} C. Pittori et al., Proceedings of
the GAMMA 2001 Symposium, to be published by AIP, and preprint (2001)
in preparation.

\end{thebibliography}
\end{document}